\documentclass[10pt,letterpaper]{article}
\usepackage{opex3}
\usepackage{cite}
\usepackage{subfigure}

\begin{document}
\title{200km Decoy-state quantum key distribution with photon polarization}
\author{Teng-Yun Chen,$^{1}$ Jian Wang,$^{1}$ Yang Liu,$^{1}$ Wen-Qi Cai,$^{1}$ Xu Wan,$^{1}$
Luo-Kan Chen,$^{1}$ Jin-Hong Wang,$^{1}$ Shu-Bin Liu,$^{1}$ Hao Liang,$^{1}$ Lin Yang$^{2,3}$,
Cheng-Zhi Peng,$^{1}$ Zeng-Bing Chen,$^{1}$ Jian-Wei Pan,$^{1}$}

\vspace{0.2cm}
\email{zbchen@ustc.edu.cn, pan@ustc.edu.cn}

\address{$^1$Hefei National Laboratory for Physical Sciences at Microscale
and Department of Modern Physics, University of Science and
Technology of China, Hefei, Anhui 230026, China\\
$^2$Department of Physics, Tsinghua University, Beijing 100084, China\\
$^3$Key Laboratory of Cryptologic Technology and Information Security, \\
Ministry of Education, Shandong University, Jinan, China}

\begin{abstract}
We demonstrate the decoy-state quantum key distribution over 200 km
with photon polarization through optical fiber, by using
super-conducting single photon detector with a repetition rate of
320 Mega Hz and a dark count rate of lower than 1 Hz. Since we have
used the polarization coding, the synchronization pulses can be run
in a low frequency. The final key rate is 14.1 Hz. The experiment
lasts for 3089 seconds with 43555 total final bits.
\end{abstract}



\section{Introduction}   Compared with the existing classical private
communication methods, quantum key distribution (QKD) is believed to
have the special advantage of unconditional security. Since Bennett
and Brasard \cite{BB84} proposed their protocol (the so called BB84
protocol) QKD has now been extensively studied both
theoretically \cite{BB84,GRTZ02,DLH06,rep,ILM,gllp,scar,cai,PNS,PNS1,
 H03, wang05, LMC05,HQph} and experimentally. The central issue
of QKD in practice is its unconditional security. The real set-up
can actually be different from the assumed ideal case in many
aspects. For example, the existing real set-ups  use an imperfect
single-photon source with a lossy channel, which is different from
the assumed ideal protocol where one can use a perfect single-photon
source. Therefore, the security is undermined by the
photon-number-splitting attack \cite{PNS,PNS1}. Fortunately, there
are a a number of methods \cite{ILM,rep, H03, wang05,
LMC05,HQph,wang07,wangapl,tdecoy,yi,pdc1,pdc2,haya,scran,kko,zei}
can be used to overcome the issue in practical QKD where we only use
the imperfect source. The ILM-GLLP proof \cite{ILM,gllp} has shown
that if we know the upper bound of fraction of the multi-photon
counts (or equivalently, the lower bound of single-photon counts)
among all raw bits, we still have a way to distill the secure final
key. Verifying such a bound is strongly non-trivial. One can
faithfully estimate such bounds with the decoy-state method
 \cite{rep, H03, wang05, LMC05,HQph}, where the intensity of pulses
are randomly changed among a few different values.

So far the decoy-state method has been extensively studied in many
experiments \cite{ron,peng,sch,yuan,tanaka,los1,los2,stucki,yama,oe,zeng2,qin,guo,lo2}
with different realization methods, including the phase coding
method \cite{ron,yuan,tanaka}, the photon polarization
method \cite{peng,sch}, in free space \cite{sch} and optical
fiber \cite{peng,yuan,tanaka}. The recently developed superconducting
detector technology can significantly improve the QKD
distance \cite{tanaka,los1,los2,stucki,yama}. So far, the most
updated record of the QKD distance is 250km \cite{stucki}, reported
by Stucki {\em et al}.
 However, the
decoy-state method is not implemented there.

Different realizations of BB84 protocol may have different technical
advantages and disadvantages. For example, polarization coding has
its advantage in {\em passively} switching the measurement basis,
while optical fiber based QKD can be directly developed for a
network secure private quantum key distribution \cite{oe}. However,
photon polarization may change significantly over long distance
transmission through an optical fiber.  Robust polarization
transmission in optical fiber has been achieved by optical
compensation \cite{peng,zeng}.

Here we report our new experimental result with photon polarization
transmitted by optical fiber, over a distance of 200km. Our result
here has kept the advantages of polarization coding and
optical-fiber based photon transmission as our earlier
result \cite{peng}, but has reached a longer secure distance by using
 superconducting detectors. Also, as shown below, together with a
superconducting detector, the polarization coding can have one more
advantage in the synchronization. In principle, the synchronization
pulses are not necessary in our system if one has a very precise
local clock at the detection side.

\section{Polarization coding, superconducting detector, and synchronization}
In a BB84 protocol \cite{BB84}, there is a sender, Alice and a
receiver, Bob. Besides sending Bob the weak coding pulses of BB84
states, Alice also sends Bob synchronization pulses as a clock, so
that Bob can know the position of each of his detected results. In
principle, the synchronization pulses do not have to be run in the
same frequency with the weak coding pulses. For example, if Bob has
a very precise clock, they don't need the high frequency
synchronization pulses in the whole protocol. However, in all
existing real set-ups, even we have such an expensive clock, the
high frequency synchronization is still necessary because they are
needed in measurement, if we use phase coding or if we use the gated
mode for the single photon detector.

Consider a system using phasing coding first. There, the signal
detection at Bob's side is done {\em actively}. Immediately before
the detecting each individual signals, an active phase shift is
taken to the signal pulse, which works as the random selection of
measurement basis. In such a system, each individual signal pulse
must be accompanied by a synchronization pulse as a clock so that
the phase shift can be done at the right time. Second, consider a
system with normal single-photon detectors, using either phase
coding or polarization coding. In this case, normally the detector
is run in a frequency of a few Mega Hertz. The detector has to be
run in a gated mode, otherwise there will be too many dark counts
and no final key can be distilled out. Such a gated mode also
requires each coding signal be accompanied by a synchronization
pulse so that the detector is gated at the right time window in
detecting each coding signal.

The situation is different if one uses polarization coding with a
superconducting detector. First, the detector's repetition rate is
so high that it can be run in the {\em always-on} mode rather than
the gated mode. Second, unlike the case of phase coding, photon
polarization detection is done {\em passively}. There is no active
operation on each signal. Therefore, in principle, in such a case,
the synchronization pulses are not necessary in the system, if one
has a very precise local clock. A very precise local clock is
technically difficult and expensive. However, the existing economic
local clock technology can run rather precisely in a short period.
This allows one to make the synchronization block by block, given
such a simple local clock at Bob's side. Say, one needs only one
synchronization pulse for a block of (many) signals.

\section{Our set-up }
Our set-up is shown schematically in Figure 1. Two sides, Alice and
Bob are linked by 200 km optical fiber. Decoy pulses of BB84
polarization states are sent to Bob through the optical fiber. They
are detected at Bob's side by superconducting single-photon
detectors.
\subsection{Source}
We use the weak coherent light as our source. The main
idea of the decoy-state method is to change intensities randomly
among different values in sending out each pulses. Here we change
the intensity of each pulses among 3 different values: 0, 0.2, 0.6,
which are called vacuum pulse, decoy-pulse, and signal pulse,
respectively. Alice's probabilities of sending a vacuum pulse, a
decoy pulse and a signal pulse are 1:1:2 . The experiment lasts for
3089 seconds.

As shown in Fig.1, the coherent light pulses in our experiment are
produced by 8 diodes which are controlled by 4-bit random numbers
random numbers. The first two bits of a random number determines
whether to produce a vacuum pulse, a decoy pulse, or a signal pulse.
In particular, if they are 00,  none of the diode sends out any
pulse, i.e., a vacuum pulse is produced; if they are 01, one of the
4 decoy diodes in the figure produces a decoy pulse; if they are 10
or 11, one of the 4 signal diodes in the figure produces a signal
pulse. The last 2 bits of the 4-bit random number decides which of
the 4 diodes is chosen to produce the pulse (If the first two bits
are not 00 ). The light intensities are controlled by an attenuator
after each diodes, 0.2 for the decoy pulse and 0.6 for the signal
pulse. Each diodes will produce only one polarization from the 4
BB84 states, i.e., the horizontal, vertical, $\pi/4$, $3\pi/4$.
Polarization maintaining beam splitters (PMBS) are used to guide
pulses from the decoy diode and the signal diode in the same
polarization block into one PMBS. Two polarization maintaining
polarization beam splitters are used to guide pulses from all diodes
to one optical fiber. The fidelity of our polarization states is
larger than $99.9\%$. The Pulses produced by the laser diodes first
passes through a polarization maintaining beam-splitter (BS),  then
a polarization beam splitter (PBS) and then combined by a single
mode BS.  These pulses are sent to Bob through the optical fiber of
a 200 km long, in side the lab.

\subsection{Detection} The circuit at Bob's side is is a standard design
of BB84 detection. Four super-conducting single-photon detectors are
used to detect signals. These detectors are made by Scontel company
in Russia. We have set the environmental temperature lower than
2.4k. The dark count rate of each detectors is smaller than 1 Hz and
the detection efficiency of 3 of them  is larger than 4\%, one of
them is larger than 3\%.

\subsection{Clock}The 40k Hz synchronization pulses are originated from the
320M Hz clock which drives a laser diode to produce synchronization
pulses. In order to observe the detect the synchronization pulses at
Bob's side which is linked to Alice by 200 km optical fiber, a
semiconductor optical amplifier is inserted at the point of 100 km,
amplifying the intensity of optical pulses by 100 before
transmission over the second half of the optical fiber. At Bob's
side, a photon-electrical detector is used to recover the
synchronization pulses. Both the synchronization pulses and the
electrical signal from the superconductor detector are sent to a
time-to-digit convertor (TDC) which works as an "economic local
clock".

\begin{figure}
\begin{center}
\includegraphics[width=95mm]{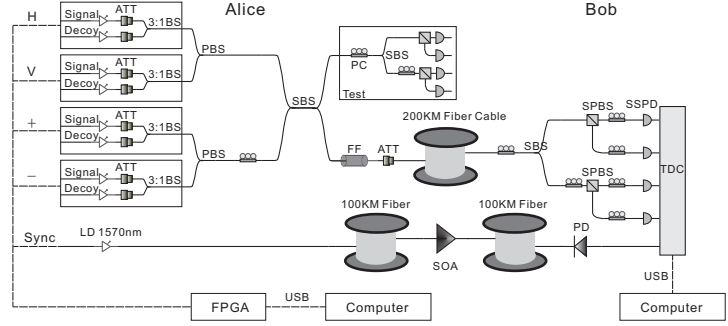}
\end{center}
\caption{\label{Fig1} Schematic diagram of our set-up. BS:
polarization maintaining beam splitter, PBS: polarization
maintaining polarization beam splitter, SBS: single mode beam
splitter, SPBS: single mode polarization beam splitter, FF: 0.2nm
bandpass filter, Att: attenuator, SOA: Semiconductor Optical
Amplifier, PD: photo detector, TDC: Time to Digital Converter. PC:
polarization controller.}
\end{figure}

\section{Calculation of the final key}
The secure final keys can be distilled with an imperfect source
given the separate theoretical results from Ref. \cite{ILM}. We
regard the upper bound of the fraction of tagged bits as those raw
bits generated by multiphoton pulses from Alice, or equivalently,
the lower bound of the fraction of untagged bits as those raw bits
generated by single-photon pulses from Alice. In Wang's 3-intensity
decoy-state protocol \cite{wang05}, Alice can randomly use 3
different intensities (average photon numbers) of each pulses ($0$,
$\mu$, $\mu'$)as the vacuum pulses, decoy pulses and signal pulses.
Alice produce the states by different intensities in
Eq.\ref{operators}.
\begin{equation}\label{operators}
\begin{array}{l}
 \rho _\mu   = e^{ - \mu } \left| 0 \right\rangle \left\langle 0 \right| + \mu e^{ - \mu } \left| 1 \right\rangle \left\langle 1 \right| + c\rho _c , \\
 \rho _{\mu '}  = e^{ - \mu '} \left| 0 \right\rangle \left\langle 0 \right| + \mu 'e^{ - \mu '} \left| 1 \right\rangle \left\langle 1 \right| + \frac{{\mu '^2 e^{ - \mu '} }}{{\mu ^2 e^{ - \mu } }}c\rho _c  + d\rho _d . \\
 \end{array}
\end{equation}
Here $c=1-e^{-\mu}-\mu e^{-\mu}$,
$\rho_c=\frac{e^{-\mu}}{c}\sum_{n=2}^{\infty}\frac{\mu^n}{n!}\left|
n \right\rangle \left\langle n \right|$,$\rho_d$ is a density
operator, and $d>0$(here we use the same notation in Ref.
 \cite{wang05}. In the protocol, Bob records all the states which he
observed his detector click. After Alice sent out all the pulses,
Bob announced his record. Then they known the number of the counts
$C_0$, $C_{\mu}$, $C_{\mu'}$ which came from different intensities
0, $\mu$, $\mu'$. $N_0$, $N_{\mu}$, $N_{\mu'}$ are the pulse numbers
of intensity $0$, $\mu$, $\mu'$ which Alice sent out. The counting
rates (the counting probability of Bob¡¯s detector whenever Alice
sends out a state) of pulses of each intensities can be calculate
$S_0=C_0/N_0$, $S_{\mu}=C_{\mu}/N_{\mu}$ and
$S_{\mu'}=C_{\mu'}/N_{\mu'}$, respectively. We denote $s_0$($s_0'$),
$s_1$($s_1'$) and $s_c$($s_c'$) for the counting rates of those
vacuum pulses, singlephoton pulses, and $\rho_c$ pulses from
$\rho_{\mu}$($\rho_{\mu'}$). Asymptotically, the values of primed
symbols here should be equal to those values of unprimed symbols.
However, in an experiment the number of samples is finite; therefore
they could be a bit different. The bound values of $s_1$, $s_1'$ can
be determined by the following joint constraints equations:
\begin{equation}\label{eq15}
\begin{array}{l}
 S_\mu   = e^{ - \mu } s_0  + \mu e^{ - \mu } s_1  + cs_c , \\
 cs_c ' \le \frac{{\mu ^2 e^{ - \mu } }}{{\mu '^2 e^{ - \mu '} }}\left( {S_{\mu '}  - \mu 'e^{ - \mu '} s'_1  - e^{ - \mu '} s'_0 } \right), \\
 \end{array}
\end{equation}
where $ s'_1  = (1 - \frac{{10e^{\mu /2} }}{{\sqrt {\mu s_1 N_\mu }
}})s_1$, $ s'_c  = (1 - \frac{{10}}{{\sqrt {s_c N_\mu  } }})s_c $,
$s_0'=(1-r_0)S_0$, $s_0=(1+r_0)S_0$, and
$r_0=\frac{10}{\sqrt{S_0N_0}}$ to obtain the worst-case results
 \cite{wang05}. Given these, one can calculate $s_1$, $s_1'$, $s_c$
numerically.

In the experiment, Alice totally transmits about N pulses to Bob.
After the transmission, Bob announces the pulse sequence numbers and
basis information of received states. Then Alice broadcasts to Bob
the actual state class information and basis information of the
corresponding pulses. Alice and Bob can calculate the experimentally
observed quantum bit error rate (QBER) values $E_{\mu}$, $E_{\mu '}$
of decoy states and signal states according to all the decoy bits
and a small fraction of the signal bits, respectively. Since we
exploit $E_{\mu}$ and $E_{\mu '}$ from finite test bit, the
statistical fluctuation should be consider to evaluate the error
rate of the remaining bit.
\begin{equation}\label{erroru}
E_{\mu'(\mu)}^U=E_{\mu'(\mu)}+10\sqrt{\frac{E_{\mu'(\mu)}}{C_{\mu'(\mu)}L_{\mu'(\mu)}^p}},
\end{equation}
$L_{\mu'(\mu)}^p$ is the proportion of the test bits for the
phase-flip test in single states (decoy state). Then we can
numerically calculate a tight lower bound of the counting rate of
single-photon $s_1'$ using Eq. (\ref{eq15}). The next step is to
estimate the fraction of single-photon $\Delta _1$ and the QBER
upper bound of single-photon $E_1$. We use
\begin{equation}\label{frac1}
\Delta _1^{\mu '}  = s_1 '\mu 'e^{ - \mu '} /S_{\mu '} ,\Delta
_1^\mu   = s_1 \mu e^{ - \mu } /S_\mu
\end{equation}
to conservatively calculate $\Delta_1$ of signal states and decoy
states, respectively \cite{wang05}. And $E_1$ of signal states and
decoy states can be estimated by the following formula:
\begin{equation}\label{error1}
E_1^{\mu'(\mu )}  = \left( {E_{\mu '(\mu )}^U  - \frac{{(1 - r_0
)S_0 e^{ - \mu '(\mu )} }}{{2S_{\mu '(\mu )} }}} \right)/\Delta
_1^{\mu '(\mu )} .
\end{equation}
Here we consider the statistical fluctuations of the vacuum states
to obtain the worst-case results.

Lastly, we can calculate the final key rates of signal states using
the following formula \cite{wang05}:
\begin{equation}\label{srate}
R_{\mu '}  = S_{\mu '} \left[ {\Delta _1^{\mu '}  - H\left( {E_{\mu
'} } \right) - \Delta _1^{\mu '} H\left( {E_1^{\mu '} } \right)}
\right].
\end{equation}
Here $H(x)=-xlog_2(x)-(1-x)log_2(1-x)$.

We consider the final key rate of the decoy states independently.
During the above calculation, we have used the worst case results in
every step for the security. Obviously, there are more economic
methods for the calculation of final key rate of the decoy states.
Here we have not considered the consumption of raw keys for the QBER
test. Now we reconsider the key rate calculation of decoy states
above. We assumed the worst case of $s_0=(1+r_0)S_0$ and
$s_0=(1-r_0)S_0$ for calculating $\Delta_1^{\mu}$ and $E_1^{\mu}$ ,
respectively. Although we do not exactly know the true value of
$s_0$, there must be one fixed value for both calculations.
Therefore we can choose every possible value in the range of
$(1-r_0)S_0\leq s_0 \leq (1+r_0)S_0$ and use it to calculate
$\Delta_1^{\mu}$, $E_1^{\mu}$ and the final key rate, and then pick
out the smallest value as the lower bound of decoy states key rate.
As the Fig. 3 in \cite{peng} , we set $s_0=(1-r_0)S_0$ to calculate
the lower bound of decoy states key rate. This economic calculation
method can obtain a more tightened value of the lower bound, which
is larger than the result using the simple calculation method above
with the two-step worst-case assumption for $s_0$ values. We can
calculate the final key rates of decoy states using the following
formula with Equ. \ref{erroru}.
\begin{equation}\label{drate}
R_{\mu}  = S_{\mu} \left[ {\Delta _1^{\mu}  - H\left( {E_{\mu} }
\right) - \Delta _1^{\mu} H\left( {E_1^{\mu} } \right)} \right].
\end{equation}
Half of the experimental data should be discarded due to the
measurement basis mismatch in the BB84 protocol. Among the remaining
half, the ratio $L_{\mu'(\mu)}^p$ are consumed for the phase-flip
test and $L_{\mu'(\mu)}^b$ are consumed for the bit-flip test. Then
we can calculate the final rate which exploit from single states and
decoy states.
\begin{equation}\label{fkey}
K_{\mu'(\mu)}=\frac{1}{2}(1-L_{\mu'(\mu)}^p-L_{\mu'(\mu)}^b)R_{\mu'(\mu)}N_{\mu'(\mu)}.
\end{equation}

In the experiment, the pulse numbers ratio of the 3 intensities 0,
$\mu$ and $\mu'$ is 1:1:2 and the intensities of signal states and
decoy states are fixed at $\mu'=0.6$ and $\mu=0.2$, respectively.
The numbers of the counts from 0, $\mu$ anf $\mu'$ are 3263, 77157
and 449467. We calculate the experimentally observed QBER values
$E_{\mu}$, $E_{\mu '}$ of decoy states and signal states are
$4.0426\%$, $1.964\%$. The experiment lasts for $T=3089$ seconds. We
use $L_{\mu'}^p=L_{\mu}^p=10\%$ for phase-flip test, and
$L_{\mu'}^b=L_{\mu}^b=5\%$ for bit-flip test.  The experimental
parameters and their corresponding values are listed in Table
\ref{t200km}.  After calculation, we obtain a final key rate of 11.8
bits/s for the signal states (intensity $\mu'=0.6$) and a final key
rate of 2.9 bits/s for the decoy states (intensity $\mu=0.2$).

\begin{table}
\centering
\caption{Experimental parameters (P) and their corresponding value of 200 km (Value) decoy-state QKD.}
\label{t200km}
  \begin{tabular}{cccccccc}
  \\ \hline
  P           & Value                 &P            & Value                   & P            & Value                  & P          & Value\\ \hline
  L           & 200 km                &$S_{\mu '}$  & $9.0941\times 10^{-7}$  &$E_{\mu'}^U$  & $0.0263$               &$E_{\mu}^U$ &  $0.0633$      \\
  $f$         & 320 MHz               &$S_{\mu}$    & $3.12225\times 10^{-7}$ &$E_1^{\mu'}$  & $0.0496$               &$E_1^{\mu}$ &  $0.0682$                  \\
  $N$         & $9.8848\times 10^{11}$&$S_0$        & $1.32041\times 10^{-8}$ &$R_{\mu'}$    & $1.7445\times 10^{-7}$ &$R_{\mu}$   &  $6.7564\times 10^{-8}$ \\
  $E_{\mu '}$ & 0.0196                &$s_1'$       & $1.2788\times 10^{-6}$  &$K_{\mu'}$    & $3.6644\times 10^4$    &$K_{\mu}$   &  $7.0960\times 10^3$     \\
  $E_{\mu }$  & 0.0404                &$s_1$        & $1.3707\times 10^{-6}$  &$K_{\mu'}/T$   & 11.8626Hz              &$K_{\mu}/T$&  2.2972Hz\\ \hline
\end{tabular}
\end{table}

\section{Concluding remarks}
In summary, with a superconducting detector, we have demonstrated a
decoy-state QKD over 200km distance in polarization coding and
optical fiber. In our experiment, the synchronization system is
significantly simplified. In principle, the frequent synchronization
pulses are not necessary in our set-up if one has a very precise
local clock.

\section*{Acknowledgments}
We acknowledge the financial support from the CAS,
the National Fundamental Research Program of China under Grant
No.2006CB921900, China Hi-Tech program grant No. 2006AA01Z420 and
the NNSFC.
\end{document}